\documentclass[prd,twocolumn,superscriptaddress,showpacs,preprintnumbers,amsmath,amssymb]{revtex4-1}

\usepackage{graphicx}
\usepackage{dcolumn}
\usepackage{bm}

\begin{document}

\preprint{}

\title{Heavy Nuclei as Thermal Insulation for Proto--Neutron Stars}

\author{Ken'ichiro Nakazato}
 \email{nakazato@artsci.kyushu-u.ac.jp}
 \affiliation{Faculty of Arts \& Science, Kyushu University, 744 Motooka, Nishi-ku, Fukuoka 819-0395, Japan
}%

\author{Hideyuki Suzuki}
 \affiliation{Faculty of Science \& Technology, Tokyo University of Science, 2641 Yamazaki, Noda, Chiba 278-8510, Japan
}%

\author{Hajime Togashi}
 \affiliation{Nishina Center for Accelerator-Based Science, RIKEN, 2-1 Hirosawa, Wako, Saitama 351-0198, Japan
}%
 \affiliation{Research Institute for Science \& Engineering, Waseda University, 3-4-1 Okubo, Shinjuku, Tokyo 169-8555, Japan
}%

\date{\today}

\begin{abstract}
A proto--neutron star (PNS) is a newly formed compact object in a core collapse supernova. In this paper, the neutrino emission from the cooling process of a PNS is investigated using two types of nuclear equation of state (EOS). It is found that the neutrino signal is mainly determined by the high-density EOS. The neutrino luminosity and mean energy are higher and the cooling time scale is longer for the softer EOS. Meanwhile, the neutrino mean energy and the cooling time scale are also affected by the low-density EOS because of the difference in the population of heavy nuclei. Heavy nuclei have a large scattering cross section with neutrinos owing to the coherent effects and act as thermal insulation near the surface of a PNS. The neutrino mean energy is higher and the cooling time scale is longer for an EOS with a large symmetry energy at low densities, namely a small density derivative coefficient of the symmetry energy, $L$.
\end{abstract}


\maketitle

\section{Introduction} \label{intro}
Radiation is a standard tool for diagnosing stellar physics not only around the surface but also inside stars. Neutrino radiation is no exception. Notably, in the case of SN1987A, neutrinos were detected for a long period of $\sim$10~s~\cite{hirata87,bionta87,alexe88}, which showed neutrino trapping inside the core of a supernova, where neutrinos undergo neutral-current interactions with nucleons~\cite{sato75}. Furthermore, heavy nuclei reside in the outer region of a supernova and near the surface of a cooling proto--neutron star (PNS), which is a nascent compact object formed as a remnant of a supernova explosion~\cite{bl86}. The neutral current cross section is enhanced by the coherent effects of heavy nuclei and is proportional to $\sim$$A^2$, with the mass number of heavy nuclei $A$~\cite{freed74}. The coherent elastic scattering of neutrinos off nuclei was recently observed in a terrestrial experiment~\cite{akimov17}. In the future, nuclear matter in a supernova and PNS will be probed through the imprint of coherent effects on the neutrino radiation~\cite{horo16}.

In this paper, we investigate the impact of heavy nuclei on the neutrino radiation from the cooling process of a PNS. The temperature of a supernova core immediately after the bounce is on the order of 10~MeV and above and there are no heavy nuclei~\cite{sumi05}. The supernova core, i.e., the PNS, is cooled by neutrino emission and, in this process, heavy nuclei appear near the surface with low densities~\cite{suzuki94}. Since the appearance of heavy nuclei is regarded as the nucleation of nuclear matter, we interpret it as a transition from a uniform phase to a nonuniform phase of inhomogeneous nuclear matter. The properties of the heavy nuclei, such as their fraction $X_{\rm A}$ and mass number $A$ at each density and temperature, are determined by nuclear interactions, as well as the nuclear equation of state (EOS).

For our study, we adopt two sets of EOS, that is, the Togashi EOS~\cite{togashi} and Shen EOS~\cite{shen}. While the Shen EOS has been used in many supernova studies, the Togashi EOS was recently constructed on the basis of variational many-body theory with the AV18 two-nucleon potential and UIX three-nucleon potential~\cite{togashi13}. Regarding uniform nuclear matter, the incompressibility, which characterizes a stiffness of EOS, of the Togashi EOS and Shen EOS are $K=245$~MeV and 281~MeV, respectively. Thus, the Togashi EOS is softer than the Shen EOS at high densities. They are different also for the properties of neutron-rich matter. At the saturation density, the symmetry energy $E_{\rm sym}$ and the density derivative coefficient of the symmetry energy $L$ are $(E_{\rm sym},L)=(30.0~{\rm MeV},35~{\rm MeV})$ for the Togashi EOS and $(E_{\rm sym},L)=(36.9~{\rm MeV},111~{\rm MeV})$ for the Shen EOS. Note that, $E_{\rm sym}$ and $L$ of the Shen EOS are somewhat higher than the current constraints~\cite{tews17}. The properties of the nonuniform phase are also different between the Togashi EOS and Shen EOS although they share the same method of dealing with inhomogeneous nuclear matter. They adopt the Thomas--Fermi approximation so as to calculate not only the EOS of the nonuniform phase but also $X_{\rm A}$ and $A$ in a self-consistent manner. Therefore, the difference in the uniform nuclear matter is reflected in the nonuniform phase~\cite{togashi,oyak07}.

Both nonuniform and uniform phases affect neutrino radiation from the cooling of a PNS~\cite{kj95,sumi95,pons99,roberts12,horo16,came17}. To extract the impacts of heavy nuclei, we prepare a third EOS connecting the Togashi EOS at high densities and the Shen EOS at low densities including the nonuniform phase, and we refer to it as the T+S EOS. The difference between the models with the Togashi EOS and T+S EOS should be attributed to heavy nuclei. In this paper, we demonstrate that heavy nuclei act as thermal insulation near the surface of a PNS, enhancing the interactions between the matter and neutrinos, and that this effect is reflected in the neutrino radiation.

\section{Setup} \label{setup}
We start our computations with the progenitor model of $15M_\odot$ in \cite{woosley95}. Utilizing the numerical code of general relativistic neutrino-radiation hydrodynamics, which solves the Boltzmann equation for neutrinos together with Lagrangian hydrodynamics under spherical symmetry \cite{sumi05}, we follow the core collapse of the progenitor until time $t=0.3$~s, which is measured from the bounce. The simulations are performed with the Togashi EOS and Shen EOS and, for both cases, the shock wave is stalled at the baryon mass coordinate of $\sim$1.47$M_\odot$ at $t=0.3$~s. From the results of these hydrodynamical simulations, the central parts of the stellar cores up to just ahead of the shock wave are extracted for use as the initial conditions of the subsequent simulations of PNS cooling, as in \cite{self13}. Incidentally, in successful supernova explosions, the PNS mass is determined self-consistently from hydrodynamical simulations \cite{hud10,fischer10,suwa14}.

In the simulations of PNS cooling, quasi-static evolutions of the PNS are solved by considering the neutrino transfer using a multigroup flux limited diffusion scheme under spherical symmetry with general relativity \cite{suzuki94}. In this method, the Boltzmann equations in the angle-integrated form are considered for $\nu_e$, $\bar \nu_e$ and $\nu_x$, where $\nu_\mu$, $\bar \nu_\mu$, $\nu_\tau$ and $\bar \nu_\tau$ are treated collectively as $\nu_x$. For the neutrino interactions, we adopt not only the standard set given by Bruenn \cite{bruenn85} but also the neutrino-positron scattering \cite{yb76} and neutrino pair processes via plasmon decay \cite{itoh89} and nucleon bremsstrahlung \cite{suzuki93}. Note that the bremsstrahlung process is important for determining the contribution of $\nu_x$. Here, the hydrostatic structure of the PNS at each time is computed by the Tolman--Oppenheimer--Volkoff equation. For simplicity, we do not take into account convection, which has been included in recent studies \cite{roberts12,horo16}.

Through the interactions between the matter and neutrinos, the profiles of the entropy and lepton number in the PNS evolve. Although we follow the core collapse with the Togashi EOS and Shen EOS individually for use as the initial conditions of the simulations of PNS cooling, we obtain similar profiles of the entropy and electron fraction as functions of the baryon mass coordinate, as in \cite{toga14}. Adopting individual initial conditions, we perform the simulations of PNS cooling with the Togashi EOS and Shen EOS. Furthermore, we evaluate the neutrino flux emitted from the cooling process of the PNS. The evolution of the PNS is computed until the central temperature drops to at least 2.2~MeV, and by then the neutrino luminosity has decreased sufficiently.

As already mentioned, we also investigate PNS cooling with the T+S EOS. For this purpose, we prepare a purely uniform nuclear matter EOS which is the same as the Togashi EOS but does not include heavy nuclei. Its property is shown in Sec.~2 of \cite{togashi} and we refer it as the Togashi' EOS. Since the baryon mass density, temperature and electron fraction are adopted as independent variables in the Togashi EOS and Shen EOS, the T+S EOS are prepared as functions of them. To obtain other variables, such as the free energy, for a given density, temperature and electron fraction, values in the Togashi' EOS and Shen EOS with the same temperature and electron fraction are interpolated in the density direction. In the T+S EOS, the Togashi' EOS for the baryon mass density $\rho_{\rm B} \ge 10^{14.3}$~g~cm$^{-3}$ and the Shen EOS for $\rho_{\rm B} \le 10^{14}$~g~cm$^{-3}$ are adopted, and the two EOSs are interpolated for 10$^{14}$~g~cm$^{-3} \le \rho_{\rm B} \le 10^{14.3}$~g~cm$^{-3}$. As a result, the T+S EOS shares the properties of heavy nuclei with the Shen EOS. In contrast, for the uniform nuclear matter with $\rho_{\rm B} \ge 10^{14.3}$~g~cm$^{-3}$, the T+S EOS is the same as the Togashi' EOS and hence Togashi EOS. The same initial condition as for the Togashi EOS is adopted for the simulation of PNS cooling with the T+S EOS.

\section{Results} \label{results}
Time evolutions of the neutrino luminosity and mean energy are shown in Figure~\ref{nulc}. The time scale of neutrino emission corresponds to that of PNS cooling. For instance, the time when the central temperature becomes sufficiently low as 2.2~MeV is $t=120.1$~s, 78.9~s and 56.6~s for Togashi EOS, T+S EOS and Shen EOS, respectively. The Togashi EOS has a longer time scale of neutrino emission than the Shen EOS. This is due to the fact that the Togashi EOS is softer and has a more compact PNS. While the baryon mass of the PNS investigated in this study is $1.47M_\odot$ for both cases, the radius of the PNS at $t=50$~s is 11.8~km for the Togashi EOS and 14.1~km for the Shen EOS. Meanwhile, the central density of the PNS is $\rho_{\rm B} = 7.73\times10^{14}$~g~cm$^{-3}$ for the Togashi EOS and $4.87\times10^{14}$~g~cm$^{-3}$ for the Shen EOS. Therefore, in the case of the Togashi EOS, the neutrino mean free path is shorter and it takes more time for neutrinos to escape from the PNS. Furthermore, the total energy emitted by neutrinos (time-integrated neutrino luminosity) is larger for Togashi EOS. This is again due to the compactness of the PNS. The energy emitted by neutrinos becomes larger for the EOS with a more compact PNS because it corresponds to the binding energy released by the cooling and shrinkage of the PNS.

According to Figure~\ref{nulc}, the neutrino luminosities for the Togashi EOS and T+S EOS are similar until $t\sim 70$~s. This means that the evolutionary histories of the PNS structure are similar for both cases. Since the main part of the PNS is a high-density region, the energy emitted as neutrinos mainly originates from there. Therefore, the evolution of the PNS is mainly determined by the high-density nuclear matter, for which the T+S EOS is the same as the Togashi EOS. In fact, at $t=50$~s, the baryon mass coordinate with $\rho_{\rm B} = 2\times10^{14}$~g~cm$^{-3}$ is $1.42M_\odot$ for both the Togashi EOS and the T+S EOS.

\clearpage
\begin{widetext}

\begin{figure}
\begin{center}
\includegraphics{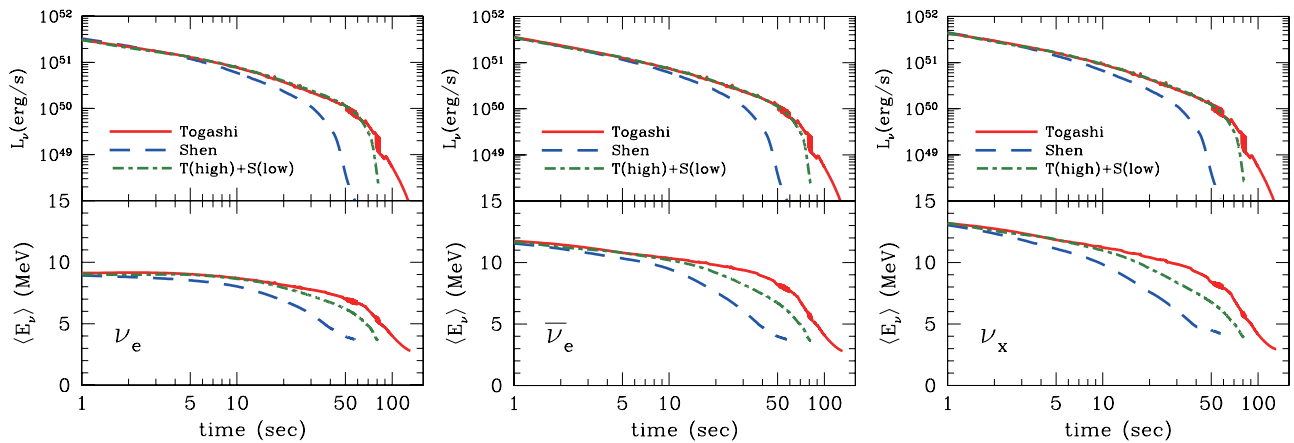}
\caption{Luminosities (upper plots) and mean energies (lower plots) of the emitted neutrinos as a function of time after the bounce. The panels correspond, from left to right, to $\nu_e$, $\bar\nu_e$ and $\nu_x$ ($=\nu_\mu$, $\nu_\tau$, $\bar\nu_\mu$, $\bar\nu_\tau$). Solid, dashed and dot-dashed lines are for the Togashi EOS, Shen EOS and T+S EOS, respectively.}
\label{nulc}
\end{center}
\end{figure}

\begin{figure}
\begin{center}
\includegraphics{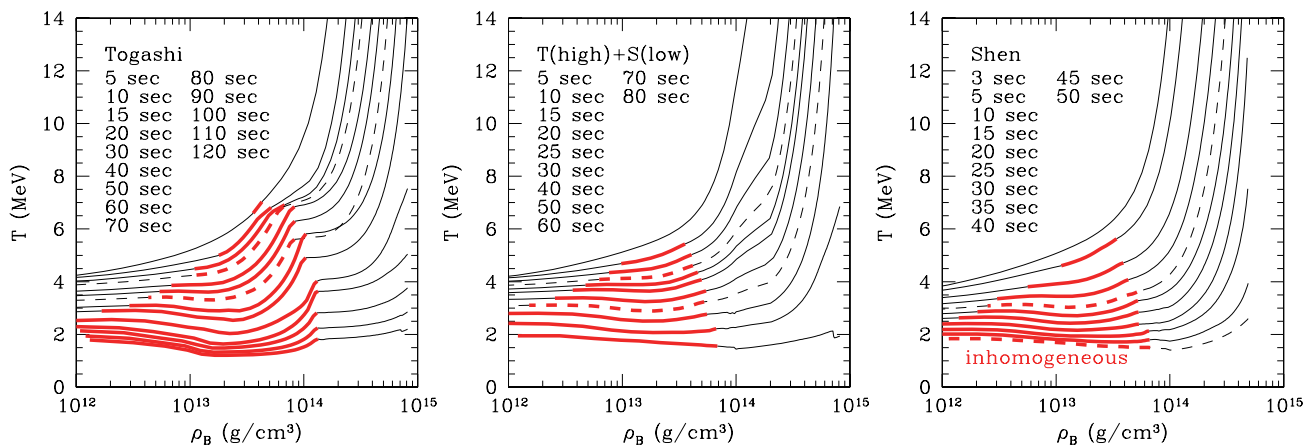}
\caption{Snapshots of PNS profiles in the baryon mass density versus temperature plane. The left, central and right panels are for the Togashi EOS, T+S EOS and Shen EOS, respectively. The lines correspond, from top to bottom, to the times listed in each panel in ascending order. The profiles at $t=20$~s and 50~s are shown in dashed lines. The nonuniform phase, where inhomogeneous nuclear matter including heavy nuclei resides, is shown by thick (red) lines and the uniform phase is shown by thin (black) lines.}
\label{phase}
\end{center}
\end{figure}

\end{widetext}

In contrast to the preceding argument, from as early as $t\sim 20$~s, a difference appears in the neutrino mean energy between the cases with the Togashi EOS and T+S EOS (Figure~\ref{nulc}). For the matter in the PNS, we show the relation between the baryon mass density and temperature at different times in Figure~\ref{phase}, in which the behaviors in the low-density region with $\rho_{\rm B} \lesssim 2\times10^{14}$~g~cm$^{-3}$ can be seen clearly. We can recognize that the case with the Togashi EOS has a higher temperature than that with the T+S EOS at any given time. Furthermore, the critical temperature and transition density between the uniform and nonuniform phases of the Togashi EOS are higher than those of the T+S EOS. For the T+S EOS, the nonuniform phase appears only in the region with $\rho_{\rm B} < 10^{14}$~g~cm$^{-3}$, where the T+S EOS and the Shen EOS are the same. The difference between the Togashi EOS and Shen EOS is described later in Sec.~\ref{dandc}.

\begin{figure}
\begin{center}
\includegraphics{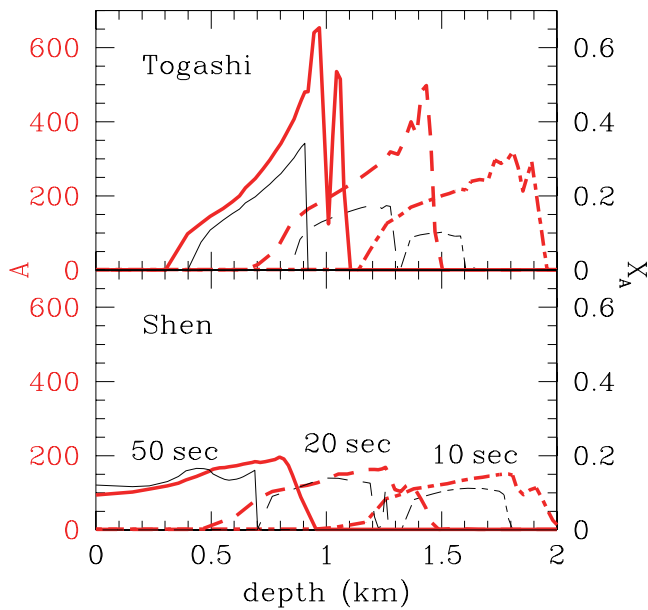}
\caption{Properties of heavy nuclei as functions of the depth from the PNS surface at 10~s (dot-dashed), 20~s (dashed) and 50~s (solid) after the bounce. Thin (black) lines represent the profiles of the fraction of heavy nuclei $X_{\rm A}$ on the right axis, and thick (red) lines represent the mass number $A$ on the left axis. The upper plots correspond to the Togashi EOS, where the radius of the PNS is 13.7~km, 12.7~km and 11.8~km at 10~s, 20~s and 50~s after the bounce, respectively. The lower plots correspond to the Shen EOS, where the radius of the PNS is 16.2~km, 15.0~km and 14.1~km at 10~s, 20~s and 50~s after the bounce, respectively.}
\label{hn}
\end{center}
\end{figure}

The properties of heavy nuclei can account for the difference in the neutrino mean energy stated above. Heavy nuclei have a large scattering cross section with neutrinos owing to the coherent effects. The cross section is proportional to $\sim$$A^2$. Meanwhile, the number density of heavy nuclei in a PNS is proportional to $\rho_{\rm B}X_{\rm A}/A$, where $X_{\rm A}$ is defined by the fraction of the number of nucleons in heavy nuclei to the total baryon number. Therefore, the local neutrino mean free path due to neutrino-nucleus scattering is proportional to $\sim$$1/(\rho_{\rm B}X_{\rm A}A)$. The values of $A$ and $X_{\rm A}$ near the surface of the PNS are shown in Figure~\ref{hn}. For the Togashi EOS, the high critical temperature hastens the appearance of heavy nuclei and the high transition density shortens the neutrino mean free path. Furthermore, as described in~\cite{togashi}, the Togashi EOS has large values of $A$ (Figure~\ref{hn}), which enhances the impact of heavy nuclei. As a result, neutrinos efficiently interact with the matter and keep the matter hot near the PNS surface. Reflecting the temperature there, the neutrino mean energy remains higher for the case with the Togashi EOS. In addition, for the Togashi EOS, hot matter near the PNS surface affects the neutrino luminosity after $t\sim 70$~s, where a difference from the T+S EOS is seen, and increases the duration of neutrino emission.

\section{Discussion and conclusion} \label{dandc}
As stated above, the interactions between the matter and neutrinos are enhanced near the surface of the PNS owing to the coherent effects of neutrino-nucleus scattering. Since neutrinos have a low energy in a PNS, they are isoenergetically scattered off heavy nuclei. It may appear strange that the isoenergetic scattering affects the state of matter, particularly the temperature. This is interpreted as follows. Owing to isoenergetic scattering, neutrinos follow a zigzag path and the probabilities of other reactions with energy exchange, such as neutrino-electron scattering, increase. Let $\lambda_{\rm th}$ be the mean free path for thermalization processes, which associates energy exchange, and $\lambda_{\rm tot}$ be the one for total reactions including isoenergetic coherent scattering. Within the mean free time for thermalizing reactions $\lambda_{\rm th}/c$ being the light velocity $c$, neutrinos react $N=\lambda_{\rm th}/\lambda_{\rm tot}$ times with the matter and propagate diffusively over the effective distance  $l_{\rm th}=\sqrt{N}\lambda_{\rm tot}=\sqrt{\lambda_{\rm tot}\lambda_{\rm th}}$. Therefore, the optical depth for thermalization at radius $r$ is given by
\begin{equation}
\tau_{\rm th}(r)=\int^{R_{\rm PNS}}_r\frac{dr^\prime}{l_{\rm th}(r^\prime)}=\int^{R_{\rm PNS}}_r\frac{dr^\prime}{\sqrt{\lambda_{\rm tot}(r^\prime)\lambda_{\rm th}(r^\prime)}},
\label{thdepth}
\end{equation}
with the PNS radius $R_{\rm PNS}$. Then the radius of the energy sphere $R_{\rm th}$ is given by $\tau_{\rm th}(R_{\rm th})=2/3$~\cite{raffelt01}. For instance, at $t=50$~s, the energy sphere of $\bar \nu_e$ with 10~MeV has radii and densities of $(R_{\rm th}, \rho_{\rm B})=($10.9~km, $6.5\times10^{13}$~g~cm$^{-3}$) for the Togashi EOS and $(R_{\rm th}, \rho_{\rm B})=($10.6~km, $1.9\times10^{14}$~g~cm$^{-3}$) for the T+S EOS. At the same time and location, the temperature is 4.9~MeV for the Togashi EOS and 4.0~MeV for the T+S EOS. This means that, for the Togashi EOS, the thermalization is achieved even in the low-density region near the PNS surface and the matter is hot.

\begin{figure}
\begin{center}
\includegraphics{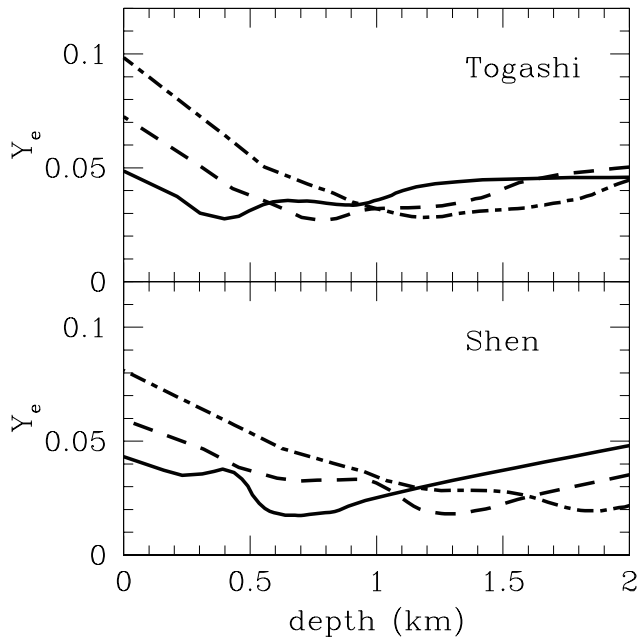}
\caption{Electron fraction as functions of the depth from the PNS surface at 10~s (dot-dashed), 20~s (dashed) and 50~s (solid) after the bounce. The upper and lower plots correspond to the Togashi EOS and Shen EOS, respectively.}
\label{ye}
\end{center}
\end{figure}

In this study, we have investigated PNS cooling with two types of EOS, the Togashi EOS and Shen EOS. The density derivative coefficient of the symmetry energy, which is one of the parameters characterizing the nuclear EOS, of the Togashi EOS ($L=35$~MeV) is smaller than that of the Shen EOS ($L=111$~MeV). Because the symmetry energy at subnuclear densities is large for the EOS with a small $L$, the matter near the PNS surface is more proton-rich for the Togashi EOS. In fact, at $t=50$~s, the electron fraction on the boundary between the uniform and nonuniform phases is 0.034 for the Togashi EOS and 0.017 for the Shen EOS, as seen in Figure~\ref{ye}. Since the creation of heavy nuclei is induced by proton clustering~\cite{oyak07}, the proton-rich matter is less stable against the creation of heavy nuclei. As a result, the temperature and density ranges of the inhomogeneous phase get wider for the Togashi EOS~\cite{togashi}, namely, (i) the critical temperature for the nonuniform phase is higher and (ii) the transition density is higher. Furthermore, for the Togashi EOS, less neutrons drip out of a nucleus in a Wigner--Seitz cell in the Thomas--Fermi model due to the larger symmetry energy in subnuclear densities. Therefore, the nucleon density of dripped matter is lower and the surface energy of nuclei becomes larger. As a result, the Coulomb energy also becomes larger because it is balanced with the surface energy. This means that the proton number $Z$ of nuclei increases. Since protons associate neutrons, (iii) the mass number of heavy nuclei $A$ is larger~\cite{oyak07}. These three factors lead to effective thermalization, keeping the matter hot near the PNS surface. Thus, through heavy nuclei, the nuclear symmetry energy is imprinted on the neutrino mean energy, which is hopefully observable for a PNS of future nearby supernova.

Beyond the aspects dealt with in this study, other possible phases are suggested for low-density nuclear matter. The pasta phase, where nuclei deform to rodlike and slablike shapes, is thought to appear just below the transition density to the uniform phase and, if this is the case, the coherent scattering of neutrinos off pasta nuclei occurs~\cite{horo04,sndy07}. In particular, pasta nuclei may appear in the density region where nuclei with the very large mass numbers reside for the Togashi EOS (Figure~\ref{hn}). In principle, to compute the scattering cross section, we should consider the full nucleon distribution reflecting the ion screening effect, which reduces the cross section~\cite{bm97,roggero17}. For pasta nuclei, the impact of coherent effects on PNS cooling was studied in~\cite{horo16} and extension of the cooling time scale was observed, as in our result. It has also been found that the pasta phase affects neutrino luminosity while the critical temperature and transition density assumed in~\cite{horo16} were higher and independent of the electron fraction. Note that the phase boundary should be determined consistently with the bulk uniform EOS for each electron fraction. In the table of the Togashi EOS, as well as that of the Shen EOS, physical quantities are given at each grid point of the density, temperature and electron fraction. Unfortunately, in this study, the discreteness of the density mesh generates numerical fluctuation of the neutrino luminosity (Figure~\ref{nulc}). Various and more precise EOS tables are desirable for future work.

In conclusion, we have performed simulations of PNS cooling with two types of EOS including heavy nuclei. We found that the neutrino luminosity is mainly determined by the high-density EOS, particularly for the early phase: the softer EOS gives a higher luminosity. On the other hand, the low-density EOS affects the cooling time scale and neutrino mean energy through heavy nuclei because they act as thermal insulation near the PNS surface owing to the coherent effect. Therefore, an EOS with a small $L$ value is implied to give a long cooling time scale and high neutrino mean energy. 

\begin{acknowledgments}
The authors are grateful to Kohsuke Sumiyoshi and Yudai Suwa for valuable comments. In this work, numerical computations were partially performed on the supercomputers at Research Center for Nuclear Physics (RCNP) in Osaka University. This work was partially supported by JSPS KAKENHI Grant Numbers JP24105008, JP26104006, JP26870615 and JP17H05203.
\end{acknowledgments}

\bibliographystyle{apsrev}
\bibliography{apssamp}

\begin{thebibliography}{10}
\expandafter\ifx\csname natexlab\endcsname\relax\def\natexlab#1{#1}\fi
\expandafter\ifx\csname bibnamefont\endcsname\relax
  \def\bibnamefont#1{#1}\fi
\expandafter\ifx\csname bibfnamefont\endcsname\relax
  \def\bibfnamefont#1{#1}\fi
\expandafter\ifx\csname citenamefont\endcsname\relax
  \def\citenamefont#1{#1}\fi
\expandafter\ifx\csname url\endcsname\relax
  \def\url#1{\texttt{#1}}\fi
\expandafter\ifx\csname urlprefix\endcsname\relax\def\urlprefix{URL }\fi
\providecommand{\bibinfo}[2]{#2}
\providecommand{\eprint}[2][]{\url{#2}}

\bibitem[{\citenamefont{Hirata et~al.}(1987)}]{hirata87}
\bibinfo{author}{\bibfnamefont{K.}~\bibnamefont{Hirata}} \bibnamefont{{\it et~al.}},
  \bibinfo{journal}{Phys.\ Rev.\ Lett.} \textbf{\bibinfo{volume}{58}},
  \bibinfo{pages}{1490} (\bibinfo{year}{1987}).

\bibitem[{\citenamefont{Bionta et~al.}(1987)}]{bionta87}
\bibinfo{author}{\bibfnamefont{R.M.}~\bibnamefont{Bionta}} \bibnamefont{{\it et~al.}},
  \bibinfo{journal}{Phys.\ Rev.\ Lett.} \textbf{\bibinfo{volume}{58}},
  \bibinfo{pages}{1494} (\bibinfo{year}{1987}).

\bibitem[{\citenamefont{Alexeyev et~al.}(1988)}]{alexe88}
\bibinfo{author}{\bibfnamefont{E.N.} \bibnamefont{Alexeyev}},
  \bibinfo{author}{\bibfnamefont{L.N.} \bibnamefont{Alexeyeva}},
  \bibinfo{author}{\bibfnamefont{I.V.} \bibnamefont{Krivosheina}},
  \bibnamefont{and} \bibinfo{author}{\bibfnamefont{V.I.} \bibnamefont{Volchenko}},
  \bibinfo{journal}{Phys.\ Lett.\ B}
  \textbf{\bibinfo{volume}{205}}, \bibinfo{pages}{209} (\bibinfo{year}{1988}).

\bibitem[{\citenamefont{Sato}(1975)}]{sato75}
\bibinfo{author}{\bibfnamefont{K.}~\bibnamefont{Sato}},
  \bibinfo{journal}{Prog.\ Theor.\ Phys.} \textbf{\bibinfo{volume}{54}},
  \bibinfo{pages}{1325} (\bibinfo{year}{1975}).

\bibitem[{\citenamefont{Burrows and Lattimer}(1986)}]{bl86}
\bibinfo{author}{\bibfnamefont{A.} \bibnamefont{Burrows}}, \bibnamefont{and}
  \bibinfo{author}{\bibfnamefont{J.M.}~\bibnamefont{Lattimer}},
  \bibinfo{journal}{Astrophys.\ J.} \textbf{\bibinfo{volume}{307}},
  \bibinfo{pages}{178} (\bibinfo{year}{1986}).

\bibitem[{\citenamefont{Freedman}(1974)}]{freed74}
\bibinfo{author}{\bibfnamefont{D.Z.}~\bibnamefont{Freedman}},
  \bibinfo{journal}{Phys.\ Rev.\ D} \textbf{\bibinfo{volume}{9}},
  \bibinfo{pages}{1389} (\bibinfo{year}{1974}).

\bibitem[{\citenamefont{Akimov et~al.}(2017)}]{akimov17}
\bibinfo{author}{\bibfnamefont{D.}~\bibnamefont{Akimov}} \bibnamefont{{\it et~al.}},
  \bibinfo{journal}{Science} \textbf{\bibinfo{volume}{357}},
  \bibinfo{pages}{1123} (\bibinfo{year}{2017}).

\bibitem[{\citenamefont{Horowitz et~al.}(2016)}]{horo16}
\bibinfo{author}{\bibfnamefont{C.J.}~\bibnamefont{Horowitz}} \bibnamefont{{\it et~al.}},
  \eprint{arXiv:1611.10226~[astro-ph.HE]}.

\bibitem[{\citenamefont{Sumiyoshi et~al.}(2005)}]{sumi05}
\bibinfo{author}{\bibfnamefont{K.}~\bibnamefont{Sumiyoshi}},
  \bibinfo{author}{\bibfnamefont{S.}~\bibnamefont{Yamada}},
  \bibinfo{author}{\bibfnamefont{H.}~\bibnamefont{Suzuki}},
  \bibinfo{author}{\bibfnamefont{H.}~\bibnamefont{Shen}},
  \bibinfo{author}{\bibfnamefont{S.}~\bibnamefont{Chiba}}, \bibnamefont{and}
  \bibinfo{author}{\bibfnamefont{H.}~\bibnamefont{Toki}},
  \bibinfo{journal}{Astrophys.\ J.} \textbf{\bibinfo{volume}{629}},
  \bibinfo{pages}{922} (\bibinfo{year}{2005}).

\bibitem[{\citenamefont{Suzuki}(1994)}]{suzuki94}
\bibinfo{author}{\bibfnamefont{H.}~\bibnamefont{Suzuki}},
  in \emph{\bibinfo{title}{Astrophysics of Neutrinos}}, edited by \bibinfo{editor}{\bibfnamefont{M.}~\bibnamefont{Fukugita}} \bibnamefont{and} \bibinfo{editor}{\bibfnamefont{A.}~\bibnamefont{Suzuki}}
  (\bibinfo{publisher}{Springer-Verlag}, Tokyo, Japan, \bibinfo{year}{1994}), p.\ \bibinfo {pages}{780}.

\bibitem[{\citenamefont{Togashi et~al.}(2017)}]{togashi}
\bibinfo{author}{\bibfnamefont{H.}~\bibnamefont{Togashi}},
  \bibinfo{author}{\bibfnamefont{K.}~\bibnamefont{Nakazato}},
  \bibinfo{author}{\bibfnamefont{Y.}~\bibnamefont{Takehara}},
  \bibinfo{author}{\bibfnamefont{S.}~\bibnamefont{Yamamuro}},
  \bibinfo{author}{\bibfnamefont{H.}~\bibnamefont{Suzuki}}, \bibnamefont{and}
  \bibinfo{author}{\bibfnamefont{M.}~\bibnamefont{Takano}},
  \bibinfo{journal}{Nucl.\ Phys.\ A} \textbf{\bibinfo{volume}{961}},
  \bibinfo{pages}{78} (\bibinfo{year}{2017}).

\bibitem[{\citenamefont{Shen et~al.}(2011)}]{shen}
\bibinfo{author}{\bibfnamefont{H.}~\bibnamefont{Shen}},
  \bibinfo{author}{\bibfnamefont{H.}~\bibnamefont{Toki}},
  \bibinfo{author}{\bibfnamefont{K.}~\bibnamefont{Oyamatsu}}, \bibnamefont{and}
  \bibinfo{author}{\bibfnamefont{K.}~\bibnamefont{Sumiyoshi}},
  \bibinfo{journal}{Nucl.\ Phys.\ A} \textbf{\bibinfo{volume}{637}},
  \bibinfo{pages}{435} (\bibinfo{year}{1998});
  \bibinfo{journal}{Prog.\ Theor.\ Phys.} \textbf{\bibinfo{volume}{100}},
  \bibinfo{pages}{1013} (\bibinfo{year}{1998});
  \bibinfo{journal}{Astrophys.\ J.\ Suppl.} \textbf{\bibinfo{volume}{197}},
  \bibinfo{pages}{20} (\bibinfo{year}{2011}).

\bibitem[{\citenamefont{Togashi and Takano}(2013)}]{togashi13}
\bibinfo{author}{\bibfnamefont{H.}~\bibnamefont{Togashi}}, \bibnamefont{and}
  \bibinfo{author}{\bibfnamefont{M.}~\bibnamefont{Takano}},
  \bibinfo{journal}{Nucl.\ Phys.\ A} \textbf{\bibinfo{volume}{902}},
  \bibinfo{pages}{53} (\bibinfo{year}{2013}).

\bibitem[{\citenamefont{Tews et~al.}(2017)}]{tews17}
\bibinfo{author}{\bibfnamefont{I.}~\bibnamefont{Tews}},
  \bibinfo{author}{\bibfnamefont{J.M.}~\bibnamefont{Lattimer}},
  \bibinfo{author}{\bibfnamefont{A.}~\bibnamefont{Ohnishi}}, \bibnamefont{and}
  \bibinfo{author}{\bibfnamefont{E.E.}~\bibnamefont{Kolomeitsev}},
  \bibinfo{journal}{Astrophys.\ J.} \textbf{\bibinfo{volume}{848}},
  \bibinfo{pages}{105} (\bibinfo{year}{2017}).

\bibitem[{\citenamefont{Oyamatsu and Iida}(2007)}]{oyak07}
\bibinfo{author}{\bibfnamefont{K.}~\bibnamefont{Oyamatsu}}, \bibnamefont{and}
  \bibinfo{author}{\bibfnamefont{K.}~\bibnamefont{Iida}},
  \bibinfo{journal}{Phys.\ Rev.\ C} \textbf{\bibinfo{volume}{75}},
  \bibinfo{pages}{015801} (\bibinfo{year}{2007}).

\bibitem[{\citenamefont{Keil and Janka}(1995)}]{kj95}
\bibinfo{author}{\bibfnamefont{W.}~\bibnamefont{Keil}}, \bibnamefont{and}
  \bibinfo{author}{\bibfnamefont{H.-T.}~\bibnamefont{Janka}},
  \bibinfo{journal}{Astron.\ Astrophys.} \textbf{\bibinfo{volume}{296}},
  \bibinfo{pages}{145} (\bibinfo{year}{1995}).

\bibitem[{\citenamefont{Sumiyoshi et~al.}(1995)}]{sumi95}
\bibinfo{author}{\bibfnamefont{K.} \bibnamefont{Sumiyoshi}},
  \bibinfo{author}{\bibfnamefont{H.} \bibnamefont{Suzuki}},
  \bibnamefont{and} \bibinfo{author}{\bibfnamefont{H.} \bibnamefont{Toki}},
  \bibinfo{journal}{Astron.\ Astrophys.} \textbf{\bibinfo{volume}{303}},
  \bibinfo{pages}{475} (\bibinfo{year}{1995}).

\bibitem[{\citenamefont{Pons et~al.}(1999)}]{pons99}
\bibinfo{author}{\bibfnamefont{J.A.}~\bibnamefont{Pons}},
  \bibinfo{author}{\bibfnamefont{S.}~\bibnamefont{Reddy}},
  \bibinfo{author}{\bibfnamefont{M.}~\bibnamefont{Prakash}},
  \bibinfo{author}{\bibfnamefont{J.M.}~\bibnamefont{Lattimer}}, \bibnamefont{and}
  \bibinfo{author}{\bibfnamefont{J.A.}~\bibnamefont{Miralles}},
  \bibinfo{journal}{Astrophys.\ J.} \textbf{\bibinfo{volume}{513}},
  \bibinfo{pages}{780} (\bibinfo{year}{1999}).

\bibitem[{\citenamefont{Roberts et~al.}(2012)}]{roberts12}
\bibinfo{author}{\bibfnamefont{L.F.}~\bibnamefont{Roberts}},
  \bibinfo{author}{\bibfnamefont{G.}~\bibnamefont{Shen}},
  \bibinfo{author}{\bibfnamefont{V.}~\bibnamefont{Cirigliano}},
  \bibinfo{author}{\bibfnamefont{J.A.}~\bibnamefont{Pons}},
  \bibinfo{author}{\bibfnamefont{S.}~\bibnamefont{Reddy}}, \bibnamefont{and}
  \bibinfo{author}{\bibfnamefont{S.E.}~\bibnamefont{Woosley}},
  \bibinfo{journal}{Phys.\ Rev.\ Lett.} \textbf{\bibinfo{volume}{108}},
  \bibinfo{pages}{061103} (\bibinfo{year}{2012}).

\bibitem[{\citenamefont{Camelio et~al.}(2017)}]{came17}
\bibinfo{author}{\bibfnamefont{G.}~\bibnamefont{Camelio}},
  \bibinfo{author}{\bibfnamefont{A.}~\bibnamefont{Lovato}},
  \bibinfo{author}{\bibfnamefont{L.}~\bibnamefont{Gualtieri}},
  \bibinfo{author}{\bibfnamefont{O.}~\bibnamefont{Benhar}},
  \bibinfo{author}{\bibfnamefont{J.A.}~\bibnamefont{Pons}}, \bibnamefont{and}
  \bibinfo{author}{\bibfnamefont{V.}~\bibnamefont{Ferrari}},
  \bibinfo{journal}{Phys.\ Rev.\ D} \textbf{\bibinfo{volume}{96}},
  \bibinfo{pages}{043015} (\bibinfo{year}{2017}).

\bibitem[{\citenamefont{Woosley and Weaver}(1995)}]{woosley95}
\bibinfo{author}{\bibfnamefont{S.E.} \bibnamefont{Woosley}}, \bibnamefont{and}
  \bibinfo{author}{\bibfnamefont{T.}~\bibnamefont{Weaver}},
  \bibinfo{journal}{Astrophys.\ J.\ Suppl.} \textbf{\bibinfo{volume}{101}},
  \bibinfo{pages}{181} (\bibinfo{year}{1995}).

\bibitem[{\citenamefont{Nakazato et~al.}(2013)}]{self13}
\bibinfo{author}{\bibfnamefont{K.}~\bibnamefont{Nakazato}},
  \bibinfo{author}{\bibfnamefont{K.}~\bibnamefont{Sumiyoshi}},
  \bibinfo{author}{\bibfnamefont{H.}~\bibnamefont{Suzuki}},
  \bibinfo{author}{\bibfnamefont{T.}~\bibnamefont{Totani}},
  \bibinfo{author}{\bibfnamefont{H.}~\bibnamefont{Umeda}}, \bibnamefont{and}
  \bibinfo{author}{\bibfnamefont{S.}~\bibnamefont{Yamada}},
  \bibinfo{journal}{Astrophys.\ J.\ Suppl.} \textbf{\bibinfo{volume}{205}},
  \bibinfo{pages}{2} (\bibinfo{year}{2013}).

\bibitem[{\citenamefont{H{\"u}depohl et~al.}(2010)}]{hud10}
\bibinfo{author}{\bibfnamefont{L.}~\bibnamefont{H{\"u}depohl}},
  \bibinfo{author}{\bibfnamefont{B.}~\bibnamefont{M{\"u}ller}},
  \bibinfo{author}{\bibfnamefont{H.-T.}~\bibnamefont{Janka}},
  \bibinfo{author}{\bibfnamefont{A.}~\bibnamefont{Marek}}, \bibnamefont{and}
  \bibinfo{author}{\bibfnamefont{G.G.}~\bibnamefont{Raffelt}},
  \bibinfo{journal}{Phys.\ Rev.\ Lett.} \textbf{\bibinfo{volume}{104}},
  \bibinfo{pages}{251101} (\bibinfo{year}{2010}).

\bibitem[{\citenamefont{Fischer et~al.}(2010)}]{fischer10}
\bibinfo{author}{\bibfnamefont{T.}~\bibnamefont{Fischer}},
  \bibinfo{author}{\bibfnamefont{S.C.}~\bibnamefont{Whitehouse}},
  \bibinfo{author}{\bibfnamefont{A.}~\bibnamefont{Mezzacappa}},
  \bibinfo{author}{\bibfnamefont{F.-K.}~\bibnamefont{Thielemann}}, \bibnamefont{and}
  \bibinfo{author}{\bibfnamefont{M.}~\bibnamefont{Liebend{\"o}rfer}},
  \bibinfo{journal}{Astron.\ Astrophys.} \textbf{\bibinfo{volume}{517}},
  \bibinfo{pages}{A80} (\bibinfo{year}{2010}).

\bibitem[{\citenamefont{Suwa}(2014)}]{suwa14}
\bibinfo{author}{\bibfnamefont{Y.}~\bibnamefont{Suwa}},
  \bibinfo{journal}{Publ.\ Astron.\ Soc.\ Jpn.} \textbf{\bibinfo{volume}{66}},
  \bibinfo{pages}{L1} (\bibinfo{year}{2014}).

\bibitem[{\citenamefont{Bruenn}(1985)}]{bruenn85}
\bibinfo{author}{\bibfnamefont{S.W.}~\bibnamefont{Bruenn}},
  \bibinfo{journal}{Astrophys.\ J.\ Suppl.} \textbf{\bibinfo{volume}{58}},
  \bibinfo{pages}{771} (\bibinfo{year}{1985}).

\bibitem[{\citenamefont{Yueh and Buchler}(1976)}]{yb76}
\bibinfo{author}{\bibfnamefont{W.R.} \bibnamefont{Yueh}}, \bibnamefont{and}
  \bibinfo{author}{\bibfnamefont{J.R.}~\bibnamefont{Buchler}},
  \bibinfo{journal}{Astrophys.\ Sp.\ Sci.} \textbf{\bibinfo{volume}{39}},
  \bibinfo{pages}{429} (\bibinfo{year}{1976}).

\bibitem[{\citenamefont{Itoh et~al.}(1989)}]{itoh89}
\bibinfo{author}{\bibfnamefont{N.}~\bibnamefont{Itoh}},
  \bibinfo{author}{\bibfnamefont{T.}~\bibnamefont{Adachi}},
  \bibinfo{author}{\bibfnamefont{M.}~\bibnamefont{Nakagawa}},
  \bibinfo{author}{\bibfnamefont{Y.}~\bibnamefont{Kohyama}}, \bibnamefont{and}
  \bibinfo{author}{\bibfnamefont{H.}~\bibnamefont{Munakata}},
  \bibinfo{journal}{Astrophys.\ J.} \textbf{\bibinfo{volume}{339}},
  \bibinfo{pages}{354} (\bibinfo{year}{1989})
  [Erratum-ibid. \textbf{360}, 741 (1990)].

\bibitem[{\citenamefont{Suzuki}(1993)}]{suzuki93}
\bibinfo{author}{\bibfnamefont{H.}~\bibnamefont{Suzuki}},
  in \emph{\bibinfo{title}{Frontiers of neutrino astrophysics}}, edited by \bibinfo{editor}{\bibfnamefont{Y.}~\bibnamefont{Suzuki}} \bibnamefont{and} \bibinfo{editor}{\bibfnamefont{K.}~\bibnamefont{Nakamura}}
  (\bibinfo{publisher}{Universal Academy Press}, Tokyo, Japan, \bibinfo{year}{1993}), p.\ \bibinfo {pages}{219}.

\bibitem[{\citenamefont{Togashi et~al.}(2014)}]{toga14}
\bibinfo{author}{\bibfnamefont{H.} \bibnamefont{Togashi}},
  \bibinfo{author}{\bibfnamefont{M.} \bibnamefont{Takano}},
  \bibinfo{author}{\bibfnamefont{K.} \bibnamefont{Sumiyoshi}},
  \bibnamefont{and} \bibinfo{author}{\bibfnamefont{K.} \bibnamefont{Nakazato}},
  \bibinfo{journal}{Prog.\ Theor.\ Exp.\ Phys.}
  \textbf{\bibinfo{volume}{2014}}, \bibinfo{pages}{023D05} (\bibinfo{year}{2014}).

\bibitem[{\citenamefont{Raffelt}(2001)}]{raffelt01}
\bibinfo{author}{\bibfnamefont{G.G.}~\bibnamefont{Raffelt}},
  \bibinfo{journal}{Astrophys.\ J.} \textbf{\bibinfo{volume}{561}},
  \bibinfo{pages}{890} (\bibinfo{year}{2001}).

\bibitem[{\citenamefont{Horowitz et~al.}(2004)}]{horo04}
\bibinfo{author}{\bibfnamefont{C.J.} \bibnamefont{Horowitz}},
  \bibinfo{author}{\bibfnamefont{M.A.} \bibnamefont{P{\'e}rez-Garc{\'{\i}}a}},
  \bibnamefont{and} \bibinfo{author}{\bibfnamefont{J.} \bibnamefont{Piekarewicz}},
  \bibinfo{journal}{Phys.\ Rev.\ C}
  \textbf{\bibinfo{volume}{69}}, \bibinfo{pages}{045804} (\bibinfo{year}{2004}).

\bibitem[{\citenamefont{Sonoda et~al.}(2007)}]{sndy07}
\bibinfo{author}{\bibfnamefont{H.}~\bibnamefont{Sonoda}},
  \bibinfo{author}{\bibfnamefont{G.}~\bibnamefont{Watanabe}},
  \bibinfo{author}{\bibfnamefont{K.}~\bibnamefont{Sato}},
  \bibinfo{author}{\bibfnamefont{T.}~\bibnamefont{Takiwaki}},
  \bibinfo{author}{\bibfnamefont{K.}~\bibnamefont{Yasuoka}}, \bibnamefont{and}
  \bibinfo{author}{\bibfnamefont{T.}~\bibnamefont{Ebisuzaki}},
  \bibinfo{journal}{Phys.\ Rev.\ C} \textbf{\bibinfo{volume}{75}},
  \bibinfo{pages}{042801} (\bibinfo{year}{2007}).

\bibitem[{\citenamefont{Bruenn and Mezzacappa}(1997)}]{bm97}
\bibinfo{author}{\bibfnamefont{S.W.}~\bibnamefont{Bruenn}}, \bibnamefont{and}
  \bibinfo{author}{\bibfnamefont{A.}~\bibnamefont{Mezzacappa}},
  \bibinfo{journal}{Phys.\ Rev.\ D} \textbf{\bibinfo{volume}{56}},
  \bibinfo{pages}{7529} (\bibinfo{year}{1997}).

\bibitem[{\citenamefont{Roggero et~al.}(2017)}]{roggero17}
\bibinfo{author}{\bibfnamefont{A.} \bibnamefont{Roggero}},
  \bibinfo{author}{\bibfnamefont{J.} \bibnamefont{Margueron}},
  \bibinfo{author}{\bibfnamefont{L.F.} \bibnamefont{Roberts}},
  \bibnamefont{and} \bibinfo{author}{\bibfnamefont{S.} \bibnamefont{Reddy}},
  \eprint{arXiv:1710.10206~[astro-ph.HE]}.

\end{thebibliography}

\end{document}